\documentclass[12pt,preprint]{aastex}

\newcommand\spitzer{\textit{Spitzer}}

\newcommand{\ltsimeq}{\raisebox{-0.6ex}{$\,\stackrel
        {\raisebox{-.2ex}{$\textstyle <$}}{\sim}\,$}}
\newcommand{\gtsimeq}{\raisebox{-0.6ex}{$\,\stackrel
        {\raisebox{-.2ex}{$\textstyle >$}}{\sim}\,$}}

\slugcomment{AJ Accepted}

\received{18 Apr 2008}
\revised{14 Jun 2008}
\accepted{20 Jun 2008}

\shorttitle{Imagery of Comet 21P}
\shortauthors{Pitticov{\'a} et al.}

\begin{document}

\title{GROUND-BASED OPTICAL AND SPITZER INFRARED IMAGING OBSERVATIONS OF
COMET 21P/GIACOBINI-ZINNER}


\author{
JANA PITTICHOV{\'A}\altaffilmark{1,2},
CHARLES E. WOODWARD\altaffilmark{3},
MICHAEL S. KELLEY\altaffilmark{4}
WILLIAM T. REACH\altaffilmark{5}
}

\altaffiltext{1}{Institute for Astronomy, University of Hawaii, 
2680 Woodlawn Drive, Honolulu, HI 96822; \\ \it{jana@ifa.hawaii.edu} }

\altaffiltext{2}{Astronomical Institute of Slovak Academy of Sciences,
D\'ubravsk\'a cesta 9, Bratislava IV, 84504, Slovakia}

\altaffiltext{3}{Department of Astronomy, School of Physics and 
Astronomy, 116 Church Street, S.~E., University of Minnesota, 
Minneapolis, MN 55455,\ \it{chelsea@astro.umn.edu} } 

\altaffiltext{4}{Department of Physics, University
of Central Florida, 4000 Central Florida Blvd., Orlando, FL
32816-2385, \ \it{msk@physics.ucf.edu} }

\altaffiltext{5}{Spitzer Science Center, MS 220-6, California Institute 
of Technology, Pasadena, CA 91125,\\ \it{reach@ipac.caltech.edu} }


\begin{abstract}

We present ground-based optical and \textit{Spitzer Space Telescope} 
infrared imaging observations of the ecliptic (Jupiter-family) comet 
21P/Giacobini-Zinner, the parent body of the Draconid 
meteor stream, during its 2005 apparition. Onset of nucleus activity 
occurred at a pre-perihelion heliocentric distance, 
$r_{h} \simeq 3.80$~AU, while post-perihelion 21P was 
dusty (peak $Af\rho = 131$~cm$^{-1}$) and active out to heliocentric 
distances $\gtsimeq 3.3$~AU following a logarithmic slope with
$r_{h}$ of $-2.04$. Coma colors, $V - R = 0.524 \pm 0.003, 
R - I = 0.487 \pm 0.004$ are redder than solar, yet comparable to 
colors derived for other Jupiter-family comets. A nucleus radius of 
$1.82 \pm 0.05$~km is derived from photometry at quiescence. \spitzer{} 
images post-perihelion exhibit an extensive coma with a prominent dust 
tail, where excess emission (over the dust continuum) in the 
4.5~\micron \ IRAC image arises from volatile gaseous CO 
and/or CO$_{2}$. No dust trail was detected 
($3\sigma$ surface brightness upper-limit of 0.3 MJy~sr$^{-1}$~pixel$^{-1}$)
along the projected velocity vector of comet 21P in the MIPS 
24~\micron{} image suggesting that the number density of trail particles 
is $\ltsimeq 7 \times 10^{-11}$~m$^{-3}$. The bolometric albedo of 
21P derived from the contemporaneous optical and \spitzer{} observations 
is $A(\theta=22\degr)=0.11$, slightly lower than values derived for 
other comets at the same phase angle.

\end{abstract}



\section{INTRODUCTION\label{intro}}

Comet nuclei formed beyond the protoplanetary disk frost line
\citep[heliocentric distances, $r_{h}$ \gtsimeq 5~AU;][]{lunine04}, 
among the giant planets and were scattered into the Kuiper Belt and
beyond into the Oort Cloud (OC). Since their formation, the interiors
and surfaces of most comets have remained at temperatures below
140 K while in ``cold storage'' in the Kuiper Belt or the OC \citep{meech04}.
Moreover, most nucleus surfaces have remained below 400 K even during 
perihelion passage. At such low temperatures, dust mineralogy remains 
stable and each comet nucleus retains a record of the minerals, ices,
and volatiles extant in the comet agglomeration zones in the early
solar system. Furthermore, comet nuclei may retain their primordial
compositional inhomogeneities so that different topographic
regions would have different compositions 
\citep[e.g.,][]{dellor07}, leading to variations
in volatile production rates as a function of nucleus rotation.
Nucleus heterogeneities are apparent in the fly-by imagery
of ecliptic comet 9P/Tempel 1. Regions of distinct topography
\citep{belton07}, and spatially distinct sites of water and CO$_{2}$
release \citep{feaga07, ahearn05} are evident, as well as
heterogeneities in surface and subsurface composition 
\citep{harker07, kado07}.

There are two general dynamical families of comets, classified by 
derived orbital elements (i.e., Tisserand parameter, $T_{J}$)
based on current observations.  Nearly-isotropic comets (NICs; $T_{J} < 2$) 
have orbits that are approximately uniformly distributed in 
inclination, and are 
derived from the OC. Ecliptic comets (EC) have orbits that are 
confined to inclinations near the ecliptic
plane, $2 \ltsimeq T_{J} \ltsimeq 3$, and originate in the Kuiper 
Belt. Resulting from frequent perihelion
passage over the 4.0 -- 4.5 Gyr period since their formation,
EC comets have become noticeably less active than OC comets,
characterized by lower gas and dust production rates. 
Multi-epoch spectral energy distributions (SEDs) of ECs, from which 
comae dust properties can be constrained, are needed to assess the possible
interrelationships between their reduced activity levels and 
dust properties.

The study of the physical properties of cometary nuclei and comae
both are equally important to our understanding of the outer solar system 
environment during the era of icy planetesimal formation, and
complement efforts to discern conditions extant in early 
protoplanetary disks during the epoch of planetesimal formation. 
We present new optical and contemporaneous \spitzer \ observations of 
comet 21P/Giacobini-Zinner obtained during its 2005 apparition obtained
as part of a larger survey of both ECs and NICs \cite[e.g.,][]{kelley06}.
Ground-based optical observations enabled us to obtain precise 
optical photometry to asses variations in dust productivity 
with heliocentric distance and to study the
the comet's near-nucleus structures, including its jets and coma.
\spitzer \ images at mid-infrared wavelengths enable the study
of the spatial distributions of volatiles and dust in the coma, as
well as facilitating investigation of comet trail properties. We describe
our observations in \S\ref{obs}, and discuss analysis of our optical
and infrared (IR) observations in \S\ref{disc}, while our conclusions are 
summarized in \S\ref{concl}. 

\section{OBSERVATIONS AND REDUCTION\label{obs}}

Comet 21P/Giacobini-Zinner was discovered by the French astronomer Michel 
Giacobini in 1900 and rediscovered two apparitions later, in 1913, by the 
German astronomer Ernst Zinner. 21P, the parent body of the Draconids 
(also know as the Giacobinids) meteor shower \citep{beech86},  
is a Jupiter-family (i.e., EC with a short-period) comet with an 
orbital period of 6.61 years and an aphelion distance just 
exterior to Jupiter's orbit.  In 1933 a spectacular Giacobini 
meteor storm was visible across Europe. The 1946 
apparition of 21P was especially noteworthy as the comet passed only 
0.26 AU from Earth in late September with a visible magnitude $\simeq 7$. 
In early October, an unexpected outburst of activity caused the comet 
to brighten to sixth magnitude. Every alternate apparition of comet 
21P presents favorable observing geometries for observers on Earth, 
with integrated coma visual magnitudes as high as 7. 21P was also the 
first comet visited by a spacecraft when the International 
Cometary Explorer (ICE) flew past at 
distance of 7,800 km on 1985 September 11 \citep{brandt88}.

\subsection{Ground-based Optical\label{obs-optuh}} 

Optical images of comet 21P/Giacobini-Zinner were obtained on the 
University of Hawai`i (UH) 2.2\,m telescope on Mauna Kea, using 
a Tektronix $2048 \times 2048$ CCD camera at the $f/10$ focus of 
the telescope (image scale of 0.219\arcsec~pixel$^{-1}$) through a 
Kron-Cousins filters set ($V$:  $\lambda_o = 5450$\AA, $\Delta\lambda$ = 
836\AA; $R$:  $\lambda_o = 6460$\AA, $\Delta\lambda$ = 1245\AA; $I$: 
$\lambda_o = 8260$\AA, $\Delta\lambda$ = 1888\AA), during multiple 
observing runs from 2004 June, 2005 October and December, and 2006 March. 
The camera read-noise was 6.0~$e^{-}$ with a gain of 1.74~$e^{-}$ per
ADU. Non-sidereal guiding at the cometary rates of motion, derived from the 
JPL/Horizons ephemeris, was used during all image observations. Specific 
observational details, including orbital geometry for 21P, filters, number 
of exposures, individual exposure times, and sky conditions  are 
summarized in Table~\ref{table:tb_obslog}.

\begin{figure}
\begin{center}
\plotone{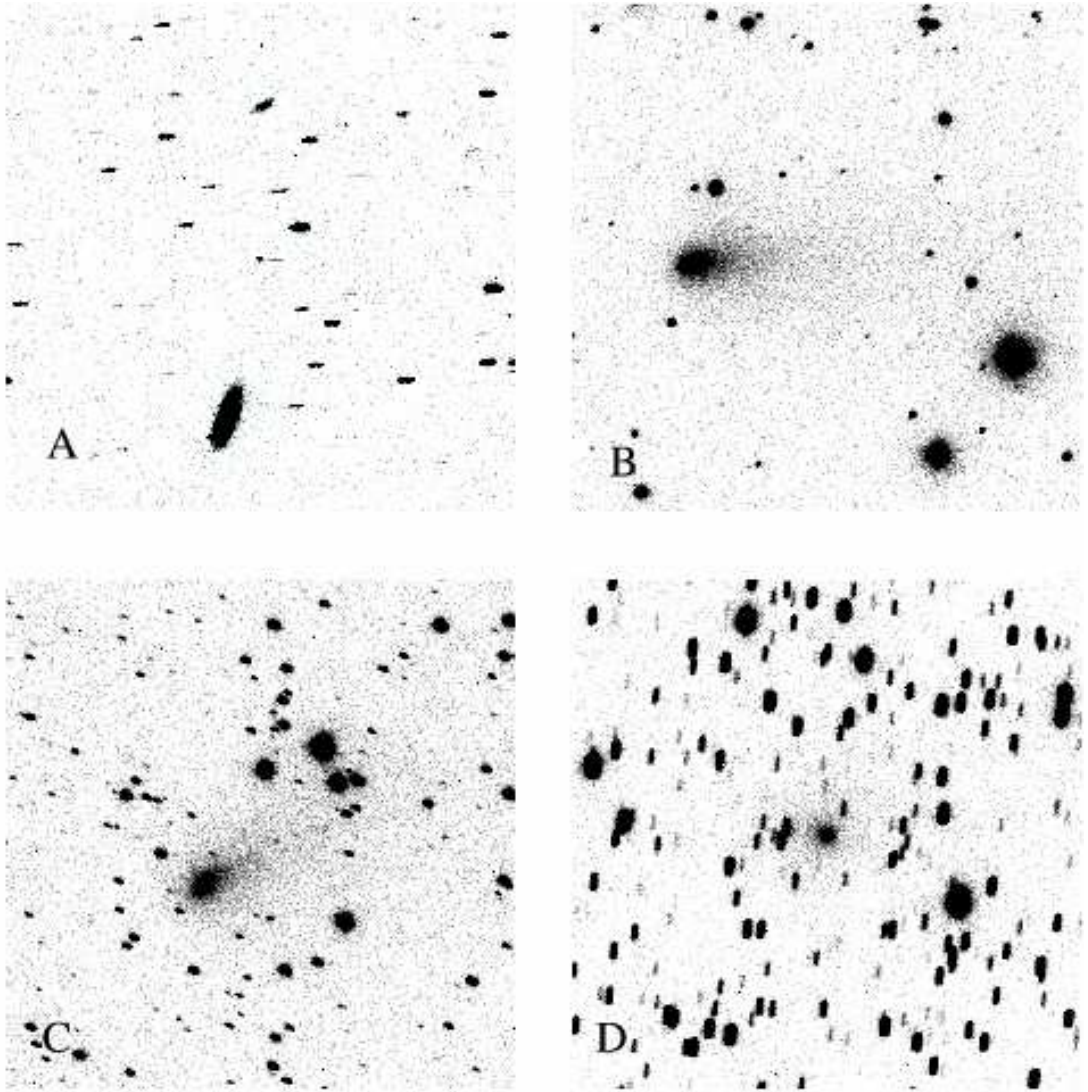}
\end{center}
\vspace{-3.0cm}
\caption{Selected optical images of comet 21P/Giacobini-Zinner obtained 
on the UH 2.2 m telescope. The 
fields of view are 180\arcsec \ $\times$ 180\arcsec \ in size. North
is up and East is to the left in all the images, while the stellar point 
sources appear elongated due to non-sidereal track rates of the telescope. 
{\it (a)} 2004 Jun 21, $r_{h}$ = 3.80 AU pre-perihelion, the comet
is marked by the ticks in the image; {\it (b)} 2005 October 
22 $r_{h}$ = 1.76 AU post-perihelion; {\it (c)} 2005 December 21 $r_{h}$ = 
2.32 AU post-perihelion; and {\it (d)} 2006 March 05, $r_{h}$ = 2.91 
AU post-perihelion.
\label{fig:opt_4panel}}
\end{figure}

The optical CCD frames were reduced with standard 
IRAF\footnote{IRAF is distributed by the National Optical Astronomy 
Observatory, which is operated by the Association of Universities for 
Research in Astronomy (AURA) under cooperative agreement with the National 
Science Foundation.} routines. All the images were 
reduced using flat field images 
taken during the evening and morning twilight sky and cleaned of bad 
pixels and cosmic rays.  The frames were calibrated with the standard 
stars of \citet{landolt92}, which were observed on each photometric night. 
Observations of typically 20 standard stars were obtained over a range of 
air-masses, and with a wide dispersion of color to fit for both extinction 
and color terms. Apparent magnitudes of twenty to thirty stars field stars 
of equal or greater brightness to the comet on each frame were measured in 
order to do relative photometry.  After correcting the measured magnitudes 
for extinction, we used the deviations of the field star magnitudes in 
each frame from their nightly average values to correct for frame-to-frame 
extinction in the comet's measured signal. Figure~\ref{fig:opt_4panel}
shows selected individual optical images for each ground-based observing run. 

\subsection{Spitzer Imaging\label{obs-ir}}

We observed comet 21P with the Infrared Array Camera
\citep[IRAC;][]{fazio04} and the Multiband Imaging Photometer for
\spitzer{} \citep[MIPS;][]{rieke04} on the \textit{Spitzer Space
Telescope} \citep{werner04}, post-perihelion as summarized in
Table~\ref{table:tb_obslog}.  The MIPS astronomical observation
request (AOR) used the 24~\micron{} array in mapping mode
to assess the existence of the
comet's dust trail (AOR key
\dataset[ADS/Sa.Spitzer#0015734784]{0015734784},
$22\arcmin\times15\arcmin$ map).  A duplicate MIPS observation was
taken 25~hr later to observe the sky background after the comet had
moved out of the frame (AOR key
\dataset[ADS/Sa.Spitzer#0015734528]{0015734528}).  The IRAC AOR
utilized the 4.5 and 8.0~\micron{} arrays to obtain complementary deep
photometry of the coma (AOR key
\dataset[ADS/Sa.Spitzer#0013820160]{0013820160},
$6.7\arcmin\times5.5\arcmin$ map).

The IRAC and MIPS observations were calibrated with \spitzer{}
pipelines S13.2.0 and S13.0.1, respectively.  The images were
mosaicked in the rest frame of the comet with the MOPEX software
\citep{makovoz05} at the native IRAC and MIPS pixel scales
(1.22\arcsec~pixel$^{-1}$ for IRAC, 2.5\arcsec~pixel$^{-1}$ for MIPS
24~\micron). The MIPS background observation was mosaicked with the same
size, scale, and orientation as the primary observation, then
subtracted from the primary.  Before mosaicking, all images were
masked to remove cosmic rays and bad pixels.  No additional
corrections were necessary.

\begin{figure}
\plotone{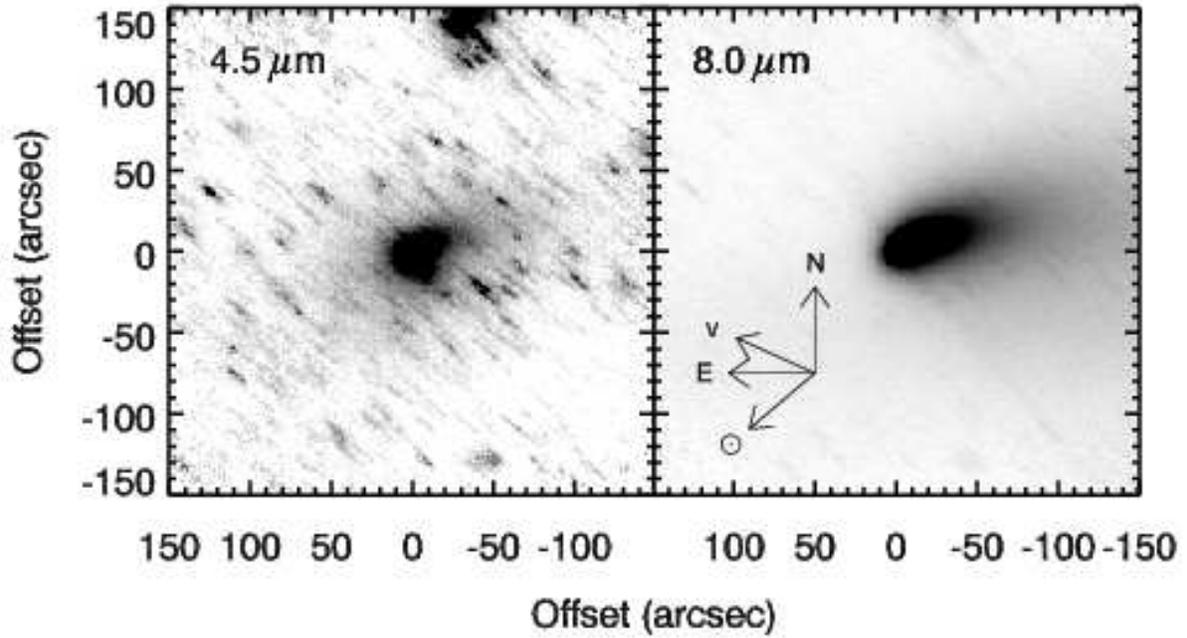}
\caption{Comet 21P/Giacobini-Zinner \spitzer{} IRAC images obtained
at $r_{h} = 2.40$~AU post-perihelion.  The
gray-scales range from 0.05 to 0.11~MJy~sr$^{-1}$ for the
4.5~\micron{} image, and 1.79 to 3.11~MJy~sr$^{-1}$ for the
8.0~\micron{} image.  \textit{Arrows} mark north, east, the
projected sun angle ($\sun$), and the projected velocity of the
comet ($v$).
\label{fig:21p-irac}}
\end{figure}

\begin{figure}
\plotone{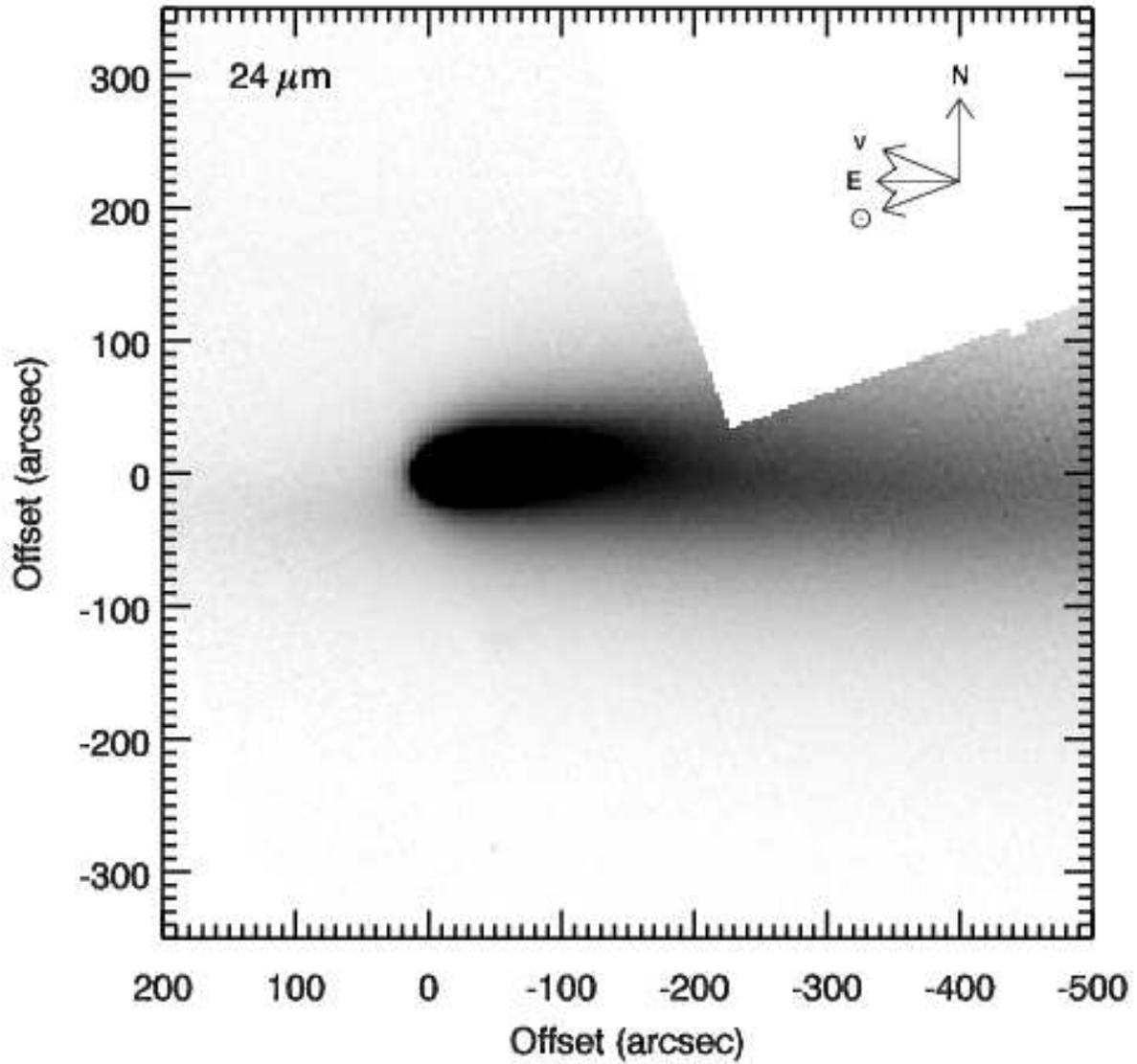}
\caption{Comet 21P/Giacobini-Zinner \spitzer{} MIPS 24~\micron{}
image obtained at $r_{h} = 2.14$~AU post-perihelion. The gray-scale 
ranges from 0.21 to 6.80~MJy~sr$^{-1}$. \textit{Arrows} mark north, 
east, the projected sun angle ($\sun$), and the projected velocity of the
comet ($v$).
\label{fig:21p-mips}}
\end{figure}

Our IRAC image of comet 21P was obtained at a post-perihelion distance
of $r_{h} = 2.40$~AU. Analysis of the secular optical light curves of 21P
by \citet{ferrin05} suggests that the nucleus turn-on time
(marking the onset of steady coma activity) occurs pre-perihelion near
3.7~AU, whereas onset of nucleus turn-off may not occur until 21P
reaches 5.4~AU. Indeed \citet{tancredi00} argue that 21P is active at
4.5~AU. The IRAC 4.5~\micron \ image, Fig.~\ref{fig:21p-irac},
shows an extended coma surrounding the nucleus. Discussion of this
extended emission is presented in \S\ref{coco2}. The \spitzer{} MIPS
24~\micron{} image is presented in Fig.~\ref{fig:21p-mips}.

\section{DISCUSSION\label{disc}}

\subsection{Optical Photometry\label{com-phot}}

The photometry derived from our optical images enable 
investigation of cometary activity with heliocentric distance, 
estimation of dust production rates \citep[$Af\rho$;][]{ahearn84}, 
determination of the coma color, and construction of 
rotational light curves.

The stellar and comet fluxes were extracted from the optical CCD
frames using the IRAF photometry routine ``PHOT'' with a circular 
aperture. This routine automatically finds the centroid
of the image within the user-specified photometry aperture, with the sky
background determined in an annulus lying immediately outside the
photometry aperture (for stellar images), or selected from an average of
many sky positions outside the extent of any coma.
Photometric apertures between 1\arcsec -- 5\arcsec \ were used
for the comet photometry with the average sky background determined
using sky annuli with an inner radii of 8.0\arcsec \ -- 10\arcsec, 
with widths of 4.0\arcsec \ -- 5.0\arcsec. Averaging many sky background
positions adjacent to the comet also enabled rejection of any bad pixels or
field stars found in individual sky annuli circumscribing the aperture used
to determine the comet surface brightness. The smallest possible aperture
that minimized contamination from extended surface brightness of the coma 
while including most of the flux from the nucleus (with respect to the
full-width half-maximum seeing determined from stellar point sources on
the frame) was 3\arcsec. A 4\arcsec \ aperture was also used for the field
stars used for differential photometry. After correcting the 
measured magnitudes for extinction, we used the deviations of the 
field stars from their nightly average magnitudes in
each frame  to correct for frame-to-frame extinction variations in the
comet's measured signal.

For the non-photometric data (Table~\ref{table:tb_obslog})
from the nights of 2004 June 21 and 22, and 2006 March 05, the comet
fields were re-imaged under photometric conditions on 2005 May 18 and 
2006 June 03 for ``boot-strap'' calibration. Multiple ($\gtsimeq 20$) field 
stars were measured on both the calibration images and the
non-photometric images and differential photometry was used to compute the
photometrically-calibrated brightness of the comet. This technique works
well for up to $\simeq 0.5$~mag of extinction.

\subsection{Heliocentric Activity\label{com-rha}}

The most direct indicator of nucleus activity is the appearance of
visible coma around a comet nucleus. Our first observations of 21P 
(Fig.~\ref{fig:opt_4panel}a) were obtained on 2004 June ($-375$~days 
before perihelion) when the comet was at a large heliocentric distance, 
$r_{h} = 3.80$~AU. No coma was evident on individual images, nor on 
the composite image created by mosaicking the 2 to 3 images obtained on 
each night to increase the signal-to-noise ratio of any diffuse coma 
emission. Absence of a detectable coma $-375$~days before perihelion is 
consistent with 1991 apparition observations of \citet{mueller92} who found 
no coma at $-367$~days before perihelion, yet a clearly detectable coma 
at $-334$ days. Our observations indicate that the heliocentric 
turn-on point for 21P nucleus activity this apparition was similar to
the behavior observed in the 1991.

The average $R$-band magnitude measured for 21P on a given night 
are summarized in Table~\ref{table:tb_mags}, column [3], as is the reduced
magnitude, column [6]. The reduced magnitude, $m(1,1,0)$, i.e., 
the observed magnitude normalized to unit heliocentric distance, $r_{h}$, and 
geocentric distance, $\Delta$, for zero phase angle, $\alpha$,  was 
derived from the relationship for active comets

\begin{equation}
m(1,1,0) = m_{R} -  2.5\,n\,log(r_{h}) - 2.5\,k\,log(\Delta) - \alpha\beta,
\label{eq:redmags}
\end{equation}

\noindent where $m_{R}$ is the $R$-band photometric magnitude measured in a
3\arcsec\ circular aperture, $\beta$ the linear phase coefficient
= 0.035~mag~deg$^{-1}$, and $n$ and $k$ are constants, taken to 
be unity and 2 respectively. Nominally, values for these two latter 
constants ($n$ being dependent on dust grain production, coma shape and 
size) are derived from modeling a given comet at a variety of 
heliocentric distances if there are sufficient observational data.  
Equation~\ref{eq:redmags} also is appropriate for expressing the brightness of 
an active comet with a spherical coma around the nucleus 
\citep[measured in a circular aperture, as described by][]{meech04}, as 
opposed to the relation used by \citet{ferrin05} which is the 
proscription for the brightness of bare nuclei (non-active stage). 
Computation of the reduced magnitude is a standard method that 
enables of photometric data of an individual comet 
obtained at a variety of heliocentric distances (different activity 
level) to be intercompared. For 21P, we obtained observations of the 
comet when there was no visible coma, when the comet exhibited a 
small coma around the nucleus, and when the comet was highly active 
displaying a significant, spatially extended coma with an elongated 
tail. To properly intercompare variations in the integrated aperture 
surface brightness, the $R$-band photometric magnitudes were 
determined by measuring the comet's brightness in a small 3\arcsec \ 
circular aperture centered on the nucleus. Use of this small 
nucleo-centric aperture minimized the coma contribution to the measured 
flux density, and allowed estimation of the nucleus brightness consistent 
with the uniform azimuthally-averaged surface brightness profile.

\begin{figure}
\plotone{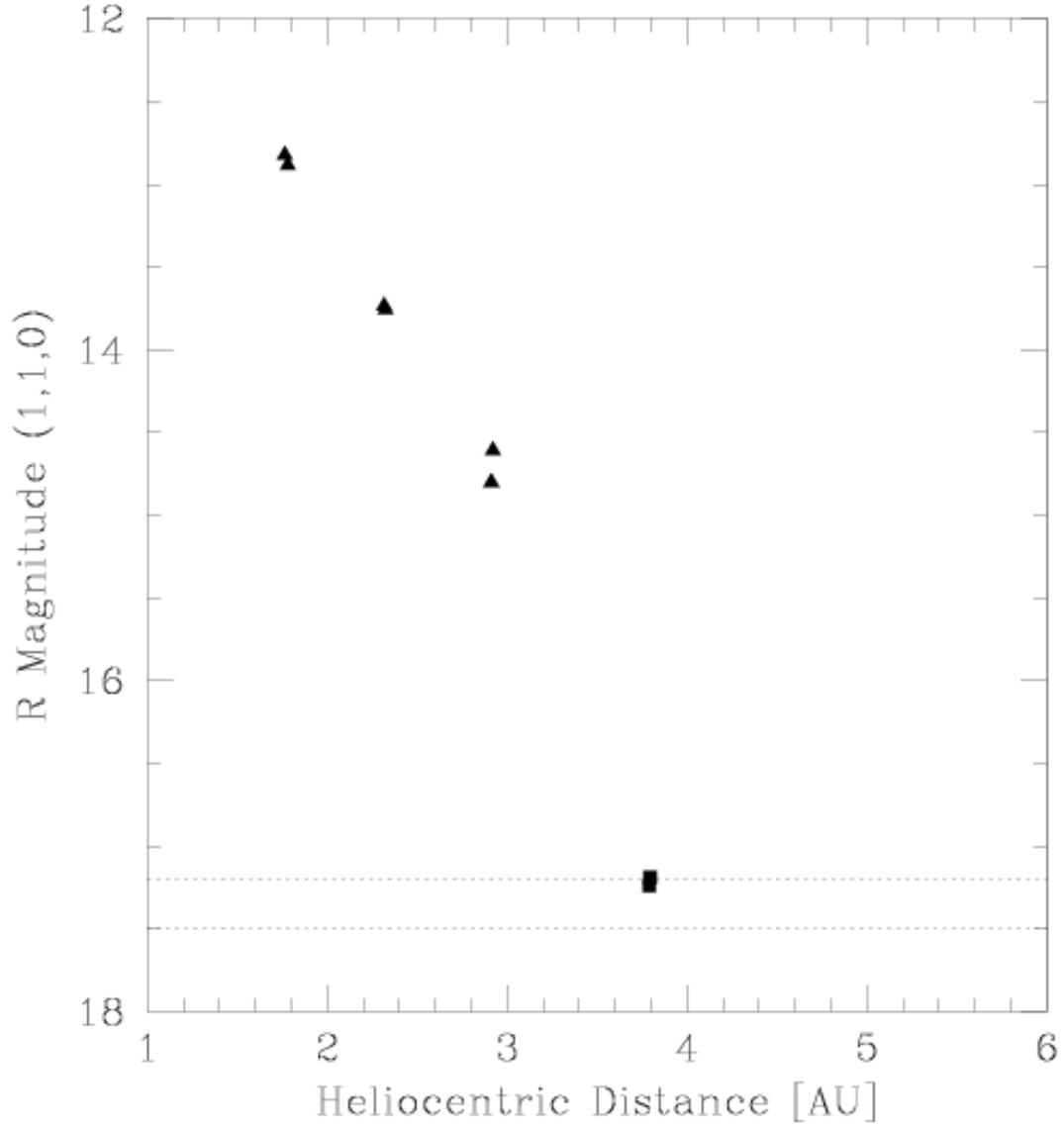}
\caption{Comet 21P/Giacobini-Zinner broadband data reduced to unit 
heliocentric distance, $r_{h}$ (AU), geocentric distance, $\Delta$ (AU),
and zero phase plotted versus $r_{h}$. The square symbols are for
pre-perihelion data and the triangular symbols are for post-perihelion 
data. The dotted horizontal lines represent the
likely brightness range for a bare nucleus, calculated from our
photometry data when no coma was seen (see Table~\ref{table:tb_mags}).
\label{fig:rm_rh}}
\end{figure}

Our observations of 21P span a range of heliocentric distance, both pre-
and post-perihelion (Table~\ref{table:tb_obslog}), thus we can examine
trends in $m(1,1,0)$ with $r_{h}$. Figure~\ref{fig:rm_rh} shows the
variation in the reduced magnitude as a function of $r_{h}$ derived from
$R$-band photometry, where the dotted horizontal lines represent the 
likely brightness range for a bare nucleus, calculated from our 
photometry data when no coma was seen. Evidently, cometary 
activity commenced near $r_{h} \simeq 3.80$~AU this apparition of 
21P, and the majority of our optical observations and all of 
our \spitzer \ images were obtained while the comet was in an active state.


\subsection{Coma Colors\label{com-cls}}


Average coma color differences ($V-R$, $R-I$) are
also listed in Table~\ref{table:tb_mags} for dates when multi-wavelength
observations were conducted. The average colors of comet 21P, $V-R =
0.524 \pm 0.003$, $R-I = 0.487 \pm 0.004$ are redder
than the respective solar colors, 0.36 and 0.28, in the Krons-Cousins system,
transforming the Johnson solar colors \citep{aq73} using the relation
of \citet{fernie83}. Band emission from gaseous 
molecular species, such as C$_{2}$ ($\lambda \approx 5520$\AA), 
NH$_{2}$ ($\lambda \approx 6335$\AA), and CN ($\lambda \approx 9180$\AA), or 
atomic O~I ($\lambda = 6300$~\AA \, $^{1}$D) potentially could 
contribute significantly to the flux density of comet comae observed 
within our $VRI$ broadband filters. However, optical spectra of 
\citet{finkhicks96} indicates that band and line emission from these species 
is relatively weak in the coma of 21P.

21P also lies along the $(R-I)$ versus $(V-R)$ color-color trend
line of Jupiter-family comet nuclei attributed by \citet{snodgrass05} to 
albedo variation (in \S~\ref{oir-albedo} we derive an dust albedo 
estimate for 21P), although we cannot quote a precise value for the nucleus 
color of 21P due to coma contamination. Broad-band colors of cometary 
nuclei are related to surface properties of the nucleus 
\citep{davidsson02} and provide a
metric to compare cometary nuclei to other small solar system bodies
and to assess which parent body populations give rise to Jupiter-family
comets. The bluer color of 21P and other Jupiter-family comets, as compared to
the mean color of Kuiper-Belt Object/Trans-Neptunian Object
populations, $V-R = 0.60 \pm 0.07$ \citep{snodgrass05}, suggests that 
cometary activity and surface processing in the inner solar system have 
altered the primitive surfaces of these nuclei agglomerated from 
materials in the Kuiper Belt \citep[e.g.,][]{jewitt02,wooden04}.

\subsection{Dust Production Rates\label{dust_afr}}


The quantity $Af\rho$ of a comet, a proxy describing the dust
production rate \citep{ahearn84}, is often used to describe 
nucleus activity vigor, and to distinguish between
dusty and inactive comet taxonomies. $Af\rho$ is given by

\begin{equation}
Af\rho = (2.467 \times 10^{19})\, \frac{r_{h}^2 \Delta}{a}
\frac{F_{obs}}{F_{\odot}},
\label{eq:afroeq}
\end{equation}

\noindent
where $A$ is the wavelength-dependent albedo of dust particles, $f$ is
the filling factor of grains (total cross-section), $\rho$ is the
linear radius of the aperture at the comet, $r_{h}$ and $\Delta$ are
heliocentric and geocentric distance (AU), $a$ is the angular diameter
of the field of view (in arcseconds), $F_{obs}$ is the observed cometary
flux and $F_{\odot}$ is the solar flux. The $Af\rho$ parameter is
typically expressed in centimeters.  By providing a metric to quantify
a comet's level of dust output (large $Af\rho$ values indicate higher
activity), one can ascertain the dust contribution to a set of 
photometric measurements and assess the heliocentric dependence of 
dust production rates in a given comet.

\begin{figure}
\plotone{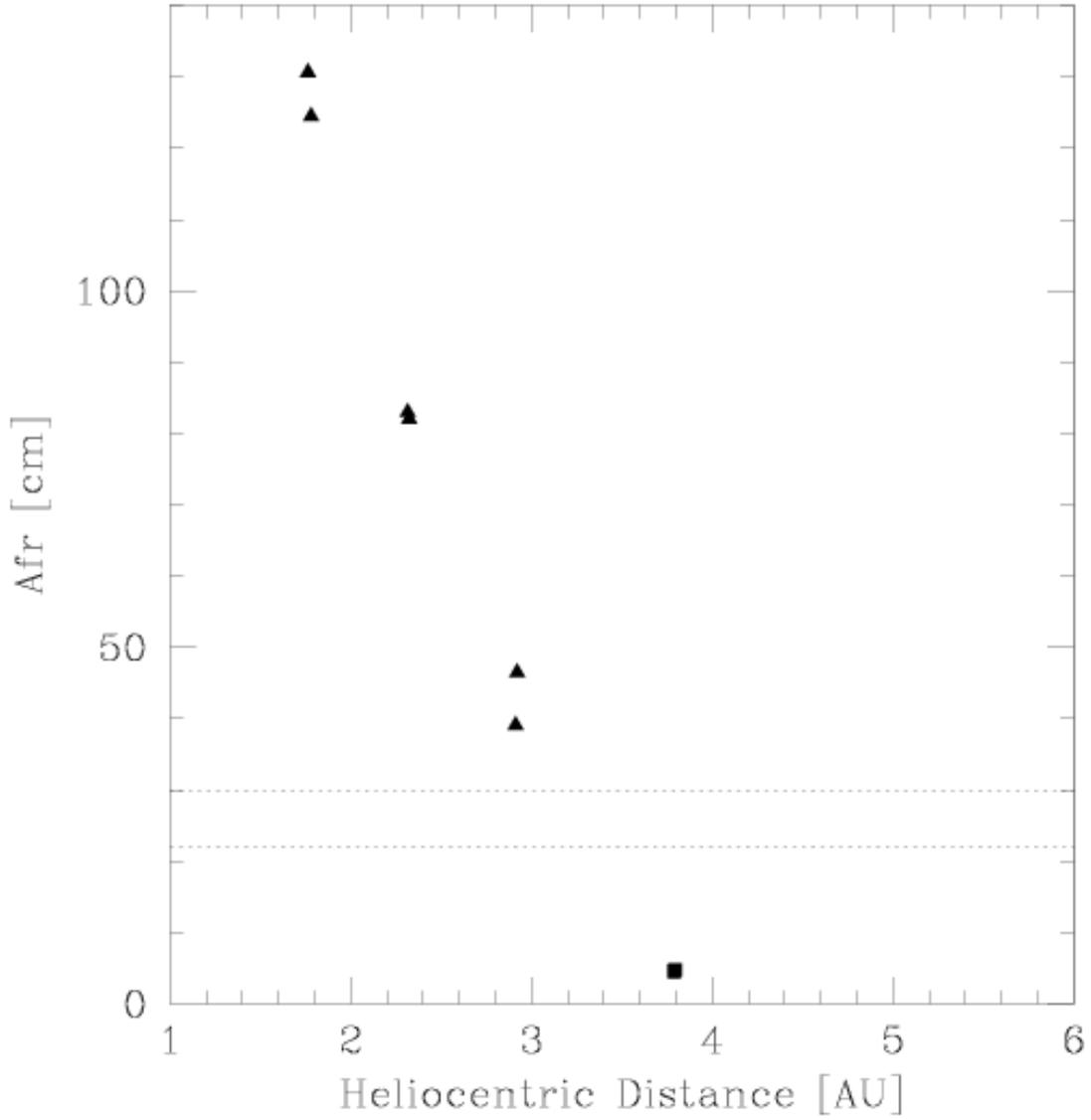}
\caption{$Af\rho$ values for comet 21P/Giacobini-Zinner as a function of
heliocentric distance, $r_{h}$ (AU), for the period between 2004 June 
through 2006 March. The square symbols are for pre-perihelion data 
and the triangular symbols denote post-perihelion data.
\label{fig:afrho_rh}}
\end{figure}

$R$-band values of $Af\rho$ for 21P derived from our photometry are
summarized in column [7] of Table~\ref{table:tb_mags} and are plotted
as a function of $r_{h}$ in Fig.~\ref{fig:afrho_rh}.  Peak dust
production rates of 131~cm$^{-1}$ occurred just post-perihelion.  We
fit the post-perihelion $Af\rho$ values with a function of the form 
$C r_{h}^k$ using the least absolute deviation method \citep{press92}.  
Our best-fit values are $C=414.4$~cm, $k=-2.04$.  The pre-perihelion
observations do not span a large enough range of $r_{h}$ to derive a
slope. Our post-perihelion $Af\rho$ slope 
($1.76 \ltsimeq r_{h} \rm{(AU)} \ltsimeq 2.92$) is comparable to the 
pre-perihelion value of $-2.08 \pm 0.15$ derived by \citet{ahearn95} based 
on observations obtained when 21P was at a $r_{h} \simeq 1.34$~AU. 
However, \citet{ahearn95} find a significantly larger value 
($k = +0.38 \pm 0.16$) when the 21P is interior to 1.86~AU 
post-perihelion suggesting that the comet's activity peaks just 
after perihelion passage.

\subsection{IR Photometry, Optical to IR SED, and the Dust
Albedo\label{oir-albedo}}

We measured the flux of comet 21P in the MIPS and IRAC images using
the aper.pro procedure of the IDL Astronomy User's Library
\citep{landsman93}.  The 2005 December IRAC, MIPS, and optical photometry 
are all measured at different heliocentric distances ($r_{h} = 2.1 - 2.4$~AU),
observing geometries ($\Delta = 1.7 - 1.9$~AU), and spatial resolutions.  
However, differences in observational circumstances are small such 
that simple scaling laws can be applied to the photometry to create a 
composite 0.5--24~\micron \  SED of comet 21P. For reasons discussed below, 
we elect to correct the IR photometry (\S~\ref{obs-ir}) to match the 
observing conditions of the optical measurements (\S~\ref{obs-optuh}).

First, we correct the IR photometry for the limited resolution of the
\spitzer{} telescope, as compared to the optical data.  The IRAC and 
MIPS images are diffraction limited and stellar sources have full-width 
half maximum $\approx 1.4$\arcsec, 2.0\arcsec, and 6\arcsec{} 
at 4.5, 8.0, and 24~\micron{} \citep{fazio04, rieke04}. The 
24~\micron{} point-spread-function (PSF)
is too broad to compare to the 3\arcsec{} $R$-band photometry in 
Table~\ref{table:tb_mags} (column [3]), and the IRAC PSF is 
marginally appropriate.  To derive the IR
photometry in an aperture size of 3741~km (corresponding to the
optical photometry, 3\arcsec{} at 1.72~AU), we measure the comet's IR
coma with increasingly larger apertures and use a curve-of-growth
analysis to estimate the fluxes at apertures comparable to or smaller
than the resolution of the \spitzer{} instruments.  We examined the
logarithmic profile of the optical data to verify that the coma slope
at moderate sized apertures ($\rho \approx 10$\arcsec) extends into the
inner-coma ($\rho \approx 3$\arcsec).  The photometry at 10\arcsec{} and
the coma logarithmic slopes are presented in Table~\ref{table:ir-photom}.

Next we correct the IRAC and MIPS photometry for the differences in
observer-comet distance.  The correction is simply 
$(\Delta_{opt} / \Delta_{IR})^2$: 0.93 for IRAC, and 1.18 for MIPS.

Finally, we scale the IRAC and MIPS photometry to match the
heliocentric distance ($r_{h} = 2.32$~AU) of the optical 
observations. There are
two effects that depend on $r_{h}$: the temperature of the dust, and the
dust production of the comet.  To correct for changes in the dust
production, we scale the photometry using the $Af\rho$ heliocentric
distance trend derived for 21P, $r_{h}^{-2.04}$ 
(\S\ref{dust_afr}). The MIPS photometry is scaled by a
factor of 0.86 and the IRAC photometry scaled by 1.08.  To correct for
the dust temperature, we scale the photometry by $B_\lambda(T_{opt}) /
B_\lambda(T_{IR})$, where $B_\lambda$ is the Planck function, and
$T_{opt}$ and $T_{IR}$ are the dust temperatures at the optical and IR
heliocentric distances, which we set to the blackbody temperature, $T
= 278\, r_{h}^{-0.5}$~K.  The scale factors are: 1.37 for
$\lambda = 4.5$~\micron, 1.20 for $\lambda = 8.0$~\micron, and 0.88 for
$\lambda = 23.7$~\micron{} for the IRAC and MIPS data respectively.

\begin{figure}
\plotone{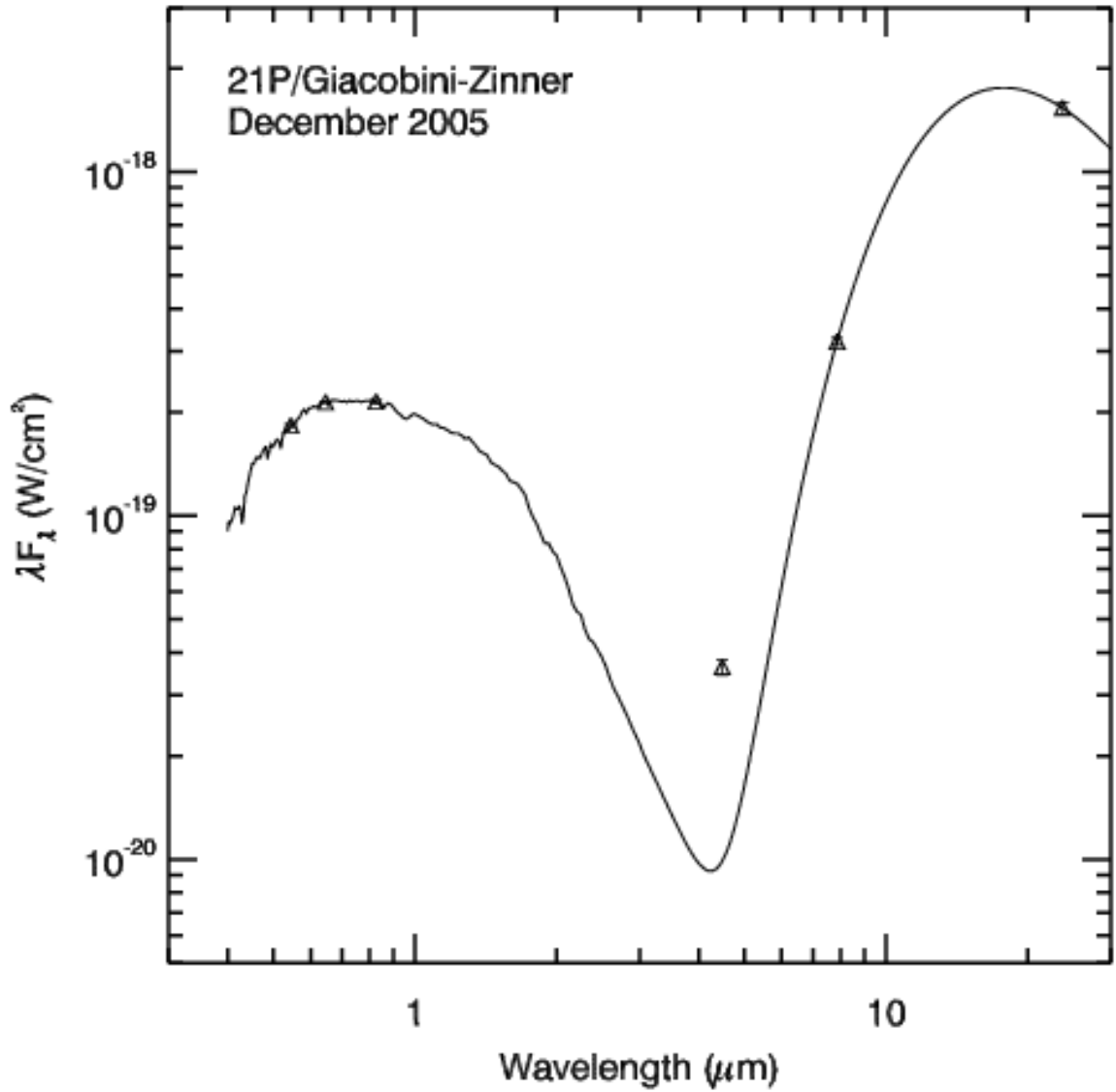}
\caption{The spectral energy distribution (SED) derived for comet
21P/Giacobini-Zinner from ground-based optical and \spitzer{} photometry
obtained in 2005 December. The smooth curve is the best fit curve
consisting of a scaled, reddened solar
spectrum to represent the scattered light at the optical wavelengths,
plus a scaled Planck function, to represent the thermal emission from
the IR wavelengths.
\label{fig:sed}}
\end{figure}

Our measured fluxes, azimuthally averaged surface brightness profile
logarithmic slopes, and the corrected photometry is presented in
Table~\ref{table:ir-photom}.  The total SED of the comet is presented
in Fig.~\ref{fig:sed}.  The optical magnitudes of
Table~\ref{table:tb_mags} were converted to flux units using the zero
magnitude flux densities of 3600~Jy ($V$-band), 3090~Jy ($R$-band),
and 2550~Jy ($I$-band).  We fit the SED with a scaled, reddened solar
spectrum to represent the scattered light at the optical wavelengths,
plus a scaled Planck function, to represent the thermal emission from
the IR wavelengths.  We exclude the gas contaminated 4.5~\micron{}
data point (\S\ref{coco2}) from the fit.  
Solar spectrum reddening (the spectral slope) is typically expressed 
as percent per 0.1~\micron{} \citep[e.g.,][]{jewitt86}.  A negative 
spectral slope represents a bluing of the spectrum.
Our observations constrain the scattered light reddening between the $V$- 
and $R$-bands, and the $R$- and $I$-bands.  Following \citet{jewitt86}, we 
express the reddening with the linear slope 
$S^{\prime}(\lambda_1,\lambda_2)$ expressed as \% per 0.1~\micron.  At
optical wavelengths, our computed slopes for comet 21P 
(see Table~\ref{table:tb_mags}) are 
$S^{\prime}(V,R)=(15\pm1)$\% per 0.1~\micron{}, 
and $S^{\prime}(R,I)=(11\pm1)$\% per 0.1~\micron{}.  The former
slope of the reflectivity gradient is in good agreement with the 
average value of $S^{\prime}(B,R)=(13\pm5)$\% per 0.1~\micron{} 
derived by \citet{jewitt86} for comets, while the latter
$S^{\prime}(R,I)$ slope value agrees with the general trend of bluer 
scattering at longer wavelengths \citep[see Fig.~4 of][]{jewitt86}.
Indeed, \citet{jewitt86} observed 21P in the near-IR and found 
$S^{\prime}(J,H)=3.5$\% per 0.1~\micron{} and $S^{\prime}(H,K)=0.3$\% per 
0.1~\micron{}.  The fit has no degrees of freedom, so we do not 
present errors or a $\chi^2$ analysis.  The best-fit dust temperature is 
206~K ($\sim 15$\% warmer than a blackbody).  We derive a bolometric 
albedo for the dust of comet 21P using Eqs.~6 and 7 from 
\citet{gehrz92}.  The bolometric albedo of 21P is $A(\theta=22\degr)=0.11$, 
slightly lower than values derived for other comets at the same phase angle 
\citep[$A(\theta = 22\degr) \approx 0.16$,][]{kolokolova05}, but not unusual.

\subsection{Nucleus Size \& Light Curves\label{com-lcs}}

The size of the equivalent spherical nucleus, $R_{N}$, can be 
estimated from the apparent mean magnitude of comet 21P measured at 
heliocentric distances when the comet is relatively inactive. 
Pre-perihelion, $m_{R} = 21.89 \pm 0.04$ (Table~\ref{table:tb_mags}) 
and $R_{N}$ can be derived from the standard relation

\begin{equation}
A_{R}R^{2}_{N} = (2.238 \times 10^{22})\, r^{2}_{h}\, \Delta^{2}\,
10^{0.4[m_{\sun} - m_{R} + \beta\alpha]} 
\label{eq:nucsiz}
\end{equation}

\noindent where $A_{R}$ is the geometric albedo, $r_{h}$~(AU) the
heliocentric distance, $\Delta$~(AU) the geocentric distance, $m_{\sun}
= -27.09$ the $R$-band magnitude of the Sun \citep{russel16}, 
$\beta$~(mag~deg$^{-1}$) the linear phase coefficient, and $\alpha$ is
the phase angle. For 21P, we adopt value of $A_{R} = 0.04$ 
\citep{lamy05,fernandez01} and  $\beta = 0.035$~mag~deg$^{-1}$ 
which is generally used for most comets with unknown $\beta$; however,
\citet{ferrin05} argue that a $\beta_{mean} = 0.046 \pm 0.013$ may
be more appropriate based on the observed phase coefficient for 10 
comets.  We find $r_{N} = 1.82 \pm 0.05$~km, commensurate with other estimates
nucleus size ranging from 1.0~km derived by \citet{tancredi00} or 
$\sim 2.1$~km value of derived from the snapshot survey of 
\citet{mueller92}, assuming $A_{R} = 0.04$.

In principle, variations in the photometric brightness derived from the 
$R$-band individual images obtained over $\approx 2-3$~hr intervals 
during the period of 2005 December 21 through 22 (Table~\ref{table:tb_mags}) 
can be used to generate a composite light curve yielding estimates of 
nucleus rotational periods. However, our composite light curve 
of 21P at this epoch, $r_{h} = 2.32$~AU exhibits little periodic variation
and we are unable to determine a rotational period. 21P was also 
very active during this epoch, with a 
significant coma contribution which could mask subtle variations 
($\pm 0.01$ mag) in nucleus flux. A rotational period of $9.5 \pm 0.2$~hr 
for 21P has been reported by \citep{leibow86}.


\subsection{Volatiles CO+CO$_{2}$\label{coco2}}

Volatiles are frozen as ices or trapped as gases in amorphous water
ice in the nuclei of comets \citep{capria02, prialnik02}.  Cometary
activity occurs when gases are released through sublimation or through
the exoergic crystallization of amorphous water ice.  Water
sublimation primarily drives vigorous activity at $r_{h} \ltsimeq 5$~AU
\citep{meech04}, but other volatiles can still be important drivers.
At 1.5~AU, \citet{feaga07} found that CO$_{2}$ gas was correlated with
comet 9P/Tempel's southern hemisphere --- the location of the strongest
dust jet \citep{farnham07} --- yet water vapor was more strongly
associated with the northern hemisphere.

Some molecular species (e.g., CO) are released not only from the
nucleus (native source), but also from within the coma
(``distributed'' or ``extended'' source). In the extensively studied,
bright OC comet C/1995~O1 (Hale-Bopp), there was evidence for distributed CO
sources, possibly arising from icy grains in the coma 
\citep{gunnaretal03a} at large $r_{h}$, 
whereas at $r_{h} \leq 1.5$~AU the strong distributed sources probably
arise from the desorption of an organic grain component 
\citep{crovis99}. Although icy grains may be the distributed source 
of the CO \citep{gunnar03, dbm02}, pure-CO ice grains do not survive 
transport into the coma because of their low sublimation 
temperatures, 20--100~K \citep{prialk04}.

The IRAC 4.5~\micron{} bandpass encompasses the CO$_{2}$ $\nu_{3}$-band
and the CO 1--0 fundamental band at 4.26 and 4.67~\micron{} respectively.
The CO and CO$_{2}$ bands are observed
as the dominate components at 3--4~\micron{} in excess of the dust 
continuum emission in \textit{ISO} spectra of comets Hale-Bopp 
and 103P/Hartley \citep{crov99a, crov99b} and in the ambient coma of 
9P/Tempel \citep{feaga07}.  After removing the dust continuum, the 
IRAC images can be used to map the CO and CO$_{2}$ spatial distributions 
in the comet coma.  In principle, the emission from the two molecules 
can be decoupled when we consider that CO and CO$_{2}$ have different
lifetimes in the coma and therefore produce different radial
distributions.  With a measured radial distribution, we can also test
for evidence of an extended source of CO or CO$_{2}$.  Emission from
other prominent volatiles, such as water \citep{woodward07}, lie
outside the bandpass of the IRAC 4.5~\micron{} filter.

\begin{figure}
\plotone{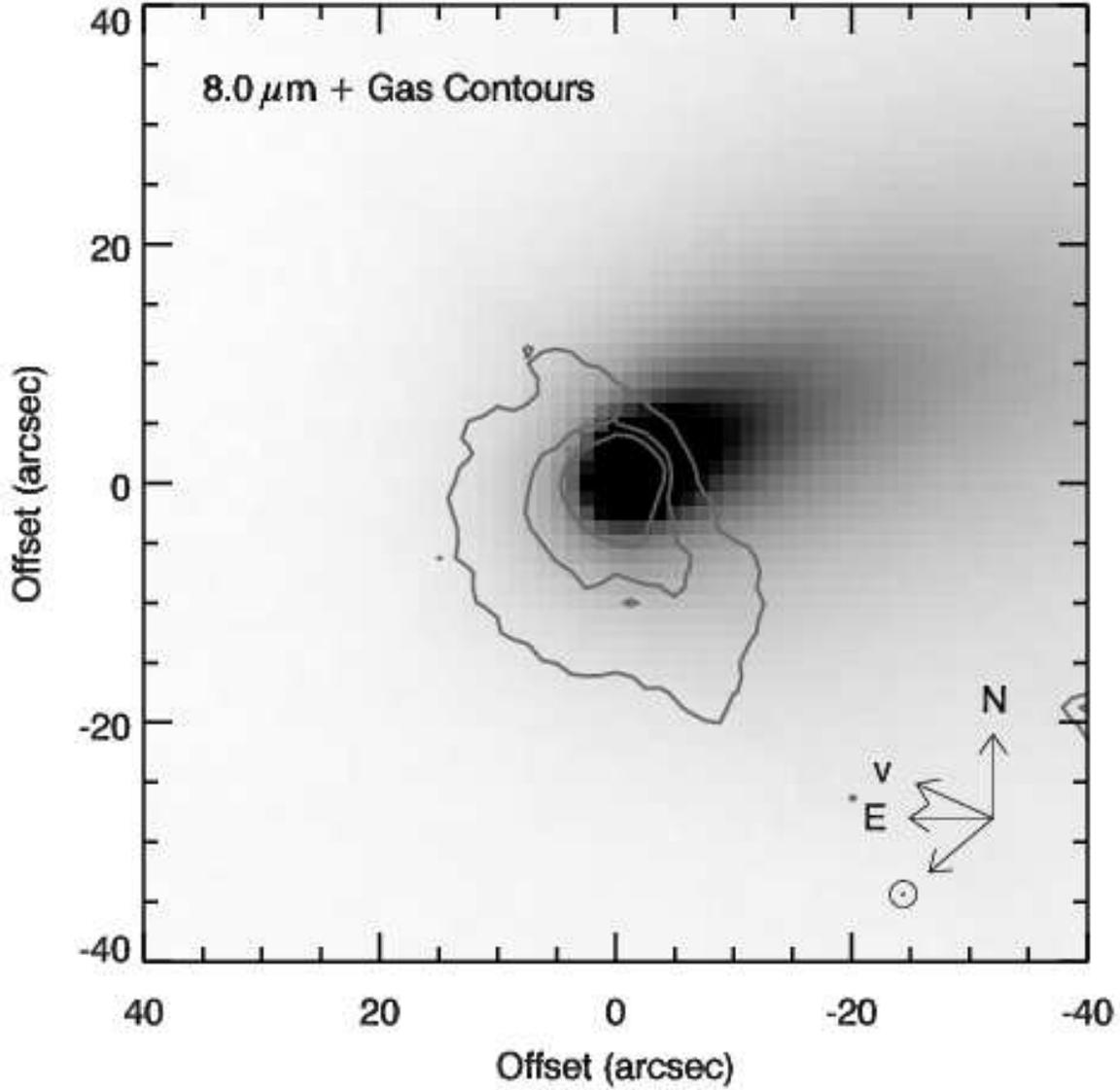}
\caption{Comet 21P/Giacobini-Zinner \spitzer{} IRAC 8.0~\micron{}
image with \textit{contours} of the excess emission (CO/CO$_{2}$ gas)
found in the 4.5~\micron{} image.  The \textit{contours} are set to
0.05, 0.10, and 0.15~MJy~sr$^{-1}$ and the gray-scale range is
1.79--7.08~MJy~sr$^{-1}$.  The surface brightness of the dust tail
peaks near the anti-sun angle, and the surface brightness of the gas
peaks along the sun angle.  \textit{Arrows} mark north, east, the
projected sun angle ($\sun$), and the projected velocity of the
comet ($v$).
\label{fig:21p-irac-gas}}
\end{figure}

The slope of the CO $+$ CO$_{2}$ emission in the \spitzer{} images at
$4-16$ pixels (corresponding to projected cometocentric distances of
$6.6 \times 10^{3}$~km to $2.6 \times 10^{4}$~km) can be determined by
subtracting a scaled IRAC 8.0~\micron{} image from the IRAC
4.5~\micron{} image (Fig.~\ref{fig:21p-irac}).  The IRAC 8.0~\micron{}
image is dominated by thermal emission from dust, whereas the
4.5~\micron{} image is a combination of emission from gas, dust
thermal emission, and sunlight scattered by dust.  The color of
sunlight scattered by comet dust varies with wavelength
\citep{jewitt86, kolokolova05}, and we cannot constrain the spectral
slope of the scattered light at 4.5~\micron{} with our data.  Rather
than estimating the 4.5~\micron{} dust flux from our best-fit SED
(\S\ref{oir-albedo}), we subtract a scaled 8.0~\micron{} image from the
4.5~\micron{} image to yield an image of the gas emission
(Fig.~\ref{fig:21p-irac-gas}). Our best scale factor is 
$0.025 \pm 0.003$, which we derived from the ratio of the 4.5~\micron{} and 
8.0~\micron{} images.  We measured the ratio along a 3 pixel wide 
rectangle positioned on the dust tail from 5 to 40 pixels from the 
nucleus. The error in the scale factor is the standard deviation of the 
pixels in the rectangle. We fit the residual surface brightness profile 
(emission from the gas) in 24\degr{} azimuthal steps. The mean surface 
brightness logarithmic slope is $-1.00$ (median $=-1.04$, error in the 
mean $=0.04$). The derived gas profile is consistent with a 
long-lived species ejected in a constant outflow from the nucleus, 
i.e, it shows no evidence for an extended source.

The lifetimes of CO and CO$_{2}$ at 2.4 AU are CO $= 89$~days and
CO$_{2} = 34$~days \citep{huebner92}.  For an outflow velocity of
1.0~km~s$^{-1}$, the characteristic length scales are $7.7 \times
10^{6}$~km (CO) and $2.9 \times 10^{6}$~km (CO$_{2}$).  In our dust
subtracted image, there appears to be emission out to 60 pixels from
the nucleus ($\simeq 1.0 \times 10^{5}$~km), but background stars make
it difficult to measure the logarithmic slope beyond 20~pixels.  Both
species are long lived on scale lengths $\leq16$ pixels even 
if the outflow velocity is
0.1~km~s$^{-1}$.  From our analysis, both CO and CO$_{2}$ are equally
viable candidates for the 4.5~\micron{} excess in the difference
image, Fig.~\ref{fig:21p-irac-gas}. For comparison, in comet
Hale-Bopp \citep{crov99a}, CO$_{2}$ was the dominant emission species
at 2.9~AU (comet 21P was observed at 2.4 AU).


We can measure the intensity of the gaseous emission and derive
upper-limits to the CO$_{2}$ and CO contributions.  We have attempted to
remove the dust from the 4.5~\micron{} image; however, emission
from the nucleus has not been removed.  Although we derived an 
effective radius from our optical observations (\S\ref{com-lcs}), 
the orientation, shape, and temperature
distribution of the nucleus is unknown at the epoch of the IRAC
images.  Rather than measuring the flux inside a circular aperture
centered on the nucleus, we avoid contamination from the nucleus by
measuring the gas coma inside an annulus with an inner-radius of
6~pixels, and an outer-radius of 8~pixels.  We convert the measured 
annular flux density, 
$(1.807\pm0.027)\times10^{27}$~erg~s$^{-1}$~cm$^{-2}$~Hz$^{-1}$, into
line emission fluxes following the IRAC prescriptions discussed by the 
\citet{idh}, where we perform a linear interpolation of values given in the 
IRAC spectral response 
tables\footnote{\url{http://ssc.caltech.edu/irac/spectral response.html}} 
near the wavelength position of our lines, assuming the residual 
emission within the effective bandpass of the IRAC 4.5~\micron{} filter 
is either entirely from either CO$_{2}$
($\lambda=4.26$~\micron) or CO ($\lambda=4.67$~\micron). We find 
$F_{CO_{2}}=(3.10\pm0.22)\times10^{-14}$~erg~s$^{-1}$~cm$^{-2}$, and
$F_{CO}=(2.86\pm0.20)\times10^{-14}$~erg~s$^{-1}$~cm$^{2}$. The errors 
include the IRAC absolute calibration uncertainty of 3\%, and a dust
subtraction uncertainty of 11\%.

The gaseous coma has an azimuthally average profile surface
brightness slope of $\rho^{-1}$ (see Table~\ref{table:ir-photom}), thus 
we can convert the flux at 6--8~pixels to a flux inside a circular 
aperture of projected 
radius 7~pixels, \, $F_{tot}=2\pi\rho^2~F_{ann}/A$, where 
$\rho=7$~pixels, $F_{ann}$ is the flux measured in the annulus, 
and $A$ is the area of the annulus in square pixels \citep[e.g.,][]{jewitt84}.  The total emission inside this 7~pixel aperture 
yields an average column density, $<N>$, in cm$^{-2}$, assuming
optically thin emission from the expression

\begin{equation}
<N>=F_{tot}~4\pi\Delta^2~\frac{\lambda}{hc}~\frac{r_h^2}{g_{band}}~\frac{1}{\pi\rho^2},
\label{eqn:moleden}
\end{equation}

\noindent where $\Delta$ is the observer-comet distance 
in cm, $\lambda$(\micron) is the
wavelength of the emission, $h$ is Planck's constant (erg~s), $c$ is the speed
of light (\micron{}~s$^{-1}$), $r_{h}$ is the heliocentric distance in
AU, $g_{band}$ is the solar fluorescence $g$-factor at 1~AU in units
of s$^{-1}$, and $\rho$ is the projected radius of the aperture, here in
units of cm.  We have adopted $g$-factors 
given by \citet{crovisier83} of $2.86\times10^{-3}$~s$^{-1}$ for the
CO$_{2}$ $\nu_3$ band, and $2.46\times10^{-4}$~s$^{-1}$ for the CO 1--0
band, noting that other cited literature values for these $g$-factors  
differ by $\lesssim$10\%.  From Eqn.~\ref{eqn:moleden}, we 
derive average column densities of: 
$(3.13\pm0.22)\times10^{11}$~cm$^{-2}$, if the emission is solely 
from CO$_{2}$ and $(3.67\pm0.27)\times10^{12}$~cm$^{-2}$, if the
emission is solely from CO. 


\citet{dbm90} measured OH radio emission profiles in spectra of 
several comets as a function of $r_{h}$, including comet
21P, to derive the expansion velocity of that species. For most 
low production rate comets ($8.0 \times 10^{28} \le 
Q_{OH}(\rm{molecules~s}^{-1}) \le 1.8 \times 10^{29}$) the expansion 
velocities are constant for $r_{h} \gtsimeq 1.4$~AU. Radio measurements 
of the OH production rates of 21P during the 1985 apparition indicate 
that $Q_{OH} \approx 3.1 \times 10^{28}$ molecules~s$^{-1}$, with little 
change in the derived expansion velocity, $v_{p} = 0.70 \pm 0.09$~km~s$^{-1}$ 
\citep[see Table 4;][]{dbm90} over a range of $r_{h}$. Given the
lack of a strong $r_{h}$ dependency of $v_{p}$ at distance greater than
1.4~AU, we will adopt $0.70 \pm 0.09$~km~s$^{-1}$ for the   
parent expansion velocity to estimate the expansion velocities of CO$_{2}$
and CO for 21P at 2.4~AU from the Sun.  With this velocity, we can derive
the production rates, $Q$, in molecules~s$^{-1}$, for each species:

\begin{equation}
  Q = <N> 2 \rho v \times 10^{10},
\end{equation}

\noindent where $\rho$ is measured in km, and $v$ in km~s$^{-1}$.  The
production rates are: $Q_{CO_{2}}\leq(5.13\pm0.75)\times10^{25}$
molecules~s$^{-1}$, and $Q_{CO}\leq(6.01\pm0.89)\times10^{26}$
molecules~s$^{-1}$.  We note that these are upper-limits to the
production rates for each species as we cannot distinguish between the
two.


\citet{mumma00} detected CO at 4.67~\micron{} in the coma of comet 21P
at $r_{h} = 1.2$~AU (pre-perihelion) in October 1998 --- the first IR
detection of this volatile in a Jupiter-family comet. However, two 
weeks later \citet{weaver99} did not detect CO at the same wavelengths.
\citet{mumma00} suggest that comet 21P has a chemically heterogeneous
nucleus, and that the CO was limited to a vent that was not
illuminated during the observations conducted by \citet{weaver99}. This
explanation seems reasonable as the comet was near perihelion when
seasonal variations of insolation are the greatest.  Our IRAC images
of comet 21P were obtained at a post-perihelion distance of $r_{h} =
2.4$~AU.  If the vent responsible for the decrease in CO production
remains inactive out to 2.4~AU, CO$_{2}$ would remain the favored
molecule responsible for the IRAC 4.5~\micron{} excess.


\subsection{Dust Trails\label{dtrls}}


Determining the cometary dust fraction of the zodiacal dust complex is
becoming increasingly important.  Recent \spitzer{} discoveries of
dust surrounding white dwarfs \citep{becklin05} suggest that mass-loss
from cometary bodies (in Kuiper Belt or Oort Cloud analogs) 
or tidal disruption of asteroidal objects are feeding
their circumstellar dust complexes \citep{su07, reach05}.  Comets lose
a large fraction of their mass in dust grains $\gtrsim100$~\micron{}
\citep{kelley08, reach07, ishiguro07, sykes92}.  These grains weakly
respond to solar radiation pressure and form distinct dynamical
structures known as dust trails \citep{sykes88}.  Trails 
typically appear to follow
their parent nucleus along the comet's projected orbit, although the
``trail'' may also lead the comet.  Dust trails are ubiquitous in
Jupiter-family comets, but remain unobserved in OC
comets.  The existence of OC dust trails is inferred from meteor
stream studies of Halley-type comets, which are derived from the OC.
For example, the Halley-type comet 55P/Tempel-Tuttle is the parent body of
the Leonid meteor stream and at least 80\% of JFCs have dust
trails \citep{reach07}.

The existence of dust trails in comets with perihelia 
near 1-2~AU has interesting astrobiological significance. 
\citet{reach07} demonstrate that comets have trail mass production 
rates that are: (1)~similar to water mass production
rates, and (2)~greater than the small grain (radii $\ltsimeq 10$~\micron) 
mass production rates.  If the
composition of the comet's coma approximates the comet interior, then
comet nuclei are mostly refractory in nature (dust-to-ice mass ratio
$> 1.0$), which limits the amount of water a comet could deliver to the
surface of a young terrestrial planet.  Comet 21P is the parent body
of the Draconid meteor stream \citep{beech86}, which can produce
exceptionally active meteor showers 
\citep[$\sim 10000$ meteors per hour;][]{jennik95}. The outburst 
activity is strongly associated
with years when the Earth and 21P closely approach each other,
suggesting that the comet is presently ejecting meteoroids.

\begin{figure}
\plotone{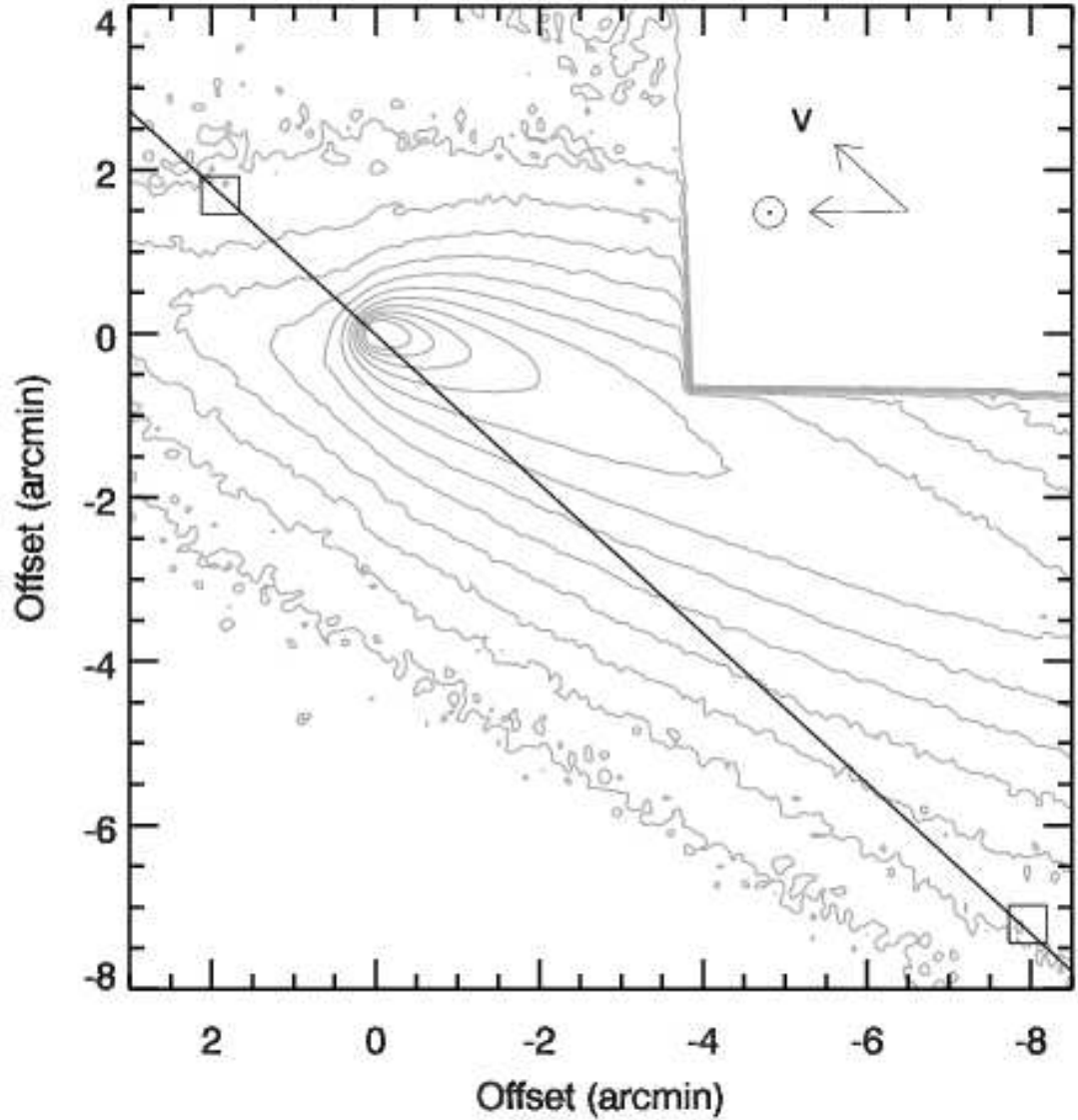}
\caption{A smoothed version of the \spitzer{} MIPS 24~\micron{} image 
($contours$) and the projected orbit of the comet 
21P/Giacobini-Zinner ({\it solid black line}).  The image has been smoothed
with a Gaussian function, 6.4\arcsec{} FWHM.  The contours are
logarithmically spaced, with each step being a 50\% increase in
surface brightness, starting at 0.3~MJy~sr$^{-1}$, and ending at
25.9~MJy~sr$^{-1}$.  $Boxes$ mark the areas where we estimated the
trail upper-limit from the standard deviation of an $11\times11$~pixel
aperture.
\label{fig:16apr08trail}}
\end{figure}

Thermal radiation from large grains in a comet trail is easily
detected at IR wavelengths from space-based telescopes \citep{reach07,
sykes92}, where the emission from warm ($\simeq 100$~K) dust peaks.
We examined our MIPS 24~\micron{} image of 21P (obtained at 2.4~AU)
for dust trail emission along the projected velocity vector of the
comet (Fig.~\ref{fig:16apr08trail}). No trail is detected with a 
$3\sigma$ surface brightness upper-limit of 0.3 MJy~sr$^{-1}$~pixel$^{-1}$.
Assuming a typical trail grain temperature of $\approx 300\,
r_{h}^{-0.5} = 193$~K \citep{sykes92}, the latter surface brightness
corresponds to an IR optical depth of $2.2 \times 10^{-9}$.  Our
upper-limit is larger than the typical comet trail observed in other
\spitzer{} MIPS images \citep{reach07}, thus it is possible that 21P
still has a dust trail.  \citet{miura07}, using deep $R$-band imaging
of 21P, do not detect the presence of brightness enhancements at the
expected position of the dust trail, and derive a upper-limit to the
number density of trail grains to be $10^{-10}$~m$^{-3}$.  Using similar
assumptions (typical grain size of 1~mm), and assuming a typical trail
thickness of $\sim10^4$~km \citep{reach07}, our MIPS optical depth
limit suggests a number density $\lesssim 7 \times
10^{-11}$~m$^{-3}$\, --- slightly lower than the optically determined
value.  \citet{miura07} also estimate the dust trail number density
from meteor shower activities.  Their lower-limit ($8 \times
10^{-15}$~m$^{-3}$) is about 4 orders of magnitude smaller than our
upper-limit.  For comparison, the trail grain number density for
81P/Wild is $\sim10^{-9}$~m$^{-3}$ \citep{ishiguro03}, and for
67P/Churyumov-Gerasimenko is $\sim 10^{-11}$~m$^{-3}$ \citep{kelley08}.
Altogether, the evidence suggests comet 21P has a dust trail fainter
than the detectable limits of the available observations.

\section{CONCLUSIONS\label{concl}}

We have presented new optical and \spitzer{} infrared observations of 
comet 21P/Giacobini-Zinner obtained during its 2005 apparition. Analysis 
of optical imagery indicates that 21P was dusty 
(peak $Af\rho = 131$~cm$^{-1}$) and active out to heliocentric distances 
$\gtsimeq 3.3$~AU following a logarithmic slope with
$r_{h}$ of $-2.04$. Onset of nucleus activity occurred at a 
pre-perihelion distance $r_{h} \simeq 3.80$~AU ($-375$~days pre-perihelion), 
similar in behavior to that observed in the 1991 apparition. The 
derived average coma colors, $V - R = 0.524 \pm 0.003, R - I = 
0.487 \pm 0.004$ are slightly redder than solar, comparable to 
colors derived for other Jupiter-family comets. Pre-perihelion 
observations during quiescence yields a nucleus radius of $1.82 \pm 0.05$~km.

\spitzer{} IRAC images obtained at $r_{h} = 2.4$~AU, post-perihelion 
exhibit an extensive coma with a prominent dust tail, where excess emission 
(over the dust continuum) in the 4.5~\micron \ image at cometocentric 
distances of $\sim 10^{4}$~km likely arises from CO$_{2}$, although a 
distributed source of CO cannot be discounted. The upper limits to
the production rates are 
$Q_{CO_{2}}\leq (5.13\pm0.75)\times10^{25}$ molecules~s$^{-1}$ and 
$Q_{CO}\leq(6.01\pm0.89)\times10^{26}$ molecules~s$^{-1}$.
The surface brightness of the gas emission is observed to peak 
along the sun angle, while the dust tail 
peaks near the anti-sunward angle. A search for dust trail emission along 
the projected velocity vector of comet 21P using our MIPS 24~\micron{} 
image ($r_{h} = 2.4$~AU), yielded no trail 
($3\sigma$ surface brightness upper-limit of 0.3 MJy~sr$^{-1}$~pixel$^{-1}$),
suggesting that the number density of trail particles (typical particle 
size $\sim 1$~mm) is $\ltsimeq 7 \times 10^{-11}$~m$^{-3}$. The bolometric
albedo of 21P derived from the contemporaneous optical and \spitzer{} 
observations is $A(\theta=22\degr)=0.11$, slightly lower than values 
derived for other comets at the same phase angle.

\acknowledgements


This work is based on observations made with the {\it Spitzer} Space
Telescope, which is operated by the Jet Propulsion Laboratory,
California Institute of Technology under a contract with NASA. Support
for this work was provided by NASA through an award issued by
JPL/Caltech.  Support for this work also was provided by NASA through
contracts 123741, 127835, and 1256406 issued by JPL/Caltech to 
the University of Minnesota.  C.E.W. also acknowledges support 
from the National Science Foundation grant AST-0706980. M.S.K. acknowledges 
support from the University of Minnesota Doctoral Dissertation Fellowship.
JP was supported, in part, by the NASA Planetary Astronomy Program, 
and by NASA through contract 1278383 issued by JPL/Caltech to the
University of Hawaii and the Scientific Grant Agency VEGA of the 
Slovak Academy of Sciences, 2/4002/04. The authors wish to thank 
Dr. Karen Meech for her insightful discussions, as well as an  
anonymous referee whose comments and suggestions improved the 
clarity of this manuscript.

{\it Facilities:} \facility{Spitzer (IRAC,MIPS)}
                  \facility{UH 2.2m (Tek2048 CCD)}


\newpage


%
%

\begin{deluxetable}{lllcccccc}
\tablewidth{0pt}
\tablecaption{OBSERVATIONAL SUMMARY LOG
\label{table:tb_obslog}}
\tablecolumns{9}
\tablehead{
\colhead{Observation}\\
\colhead{Date}
& \colhead{}
& \colhead{$r$}
& \colhead{$\Delta$}
& \colhead{$\alpha$\tablenotemark{a}}
& \colhead{FWHM\tablenotemark{b}}
& \colhead{Filter}
& \colhead{Integration}
& \colhead{}\\
\colhead{(UT)}
& \colhead{Telescope}
& \colhead{(AU)}
& \colhead{(AU)}
& \colhead{(deg)}
& \colhead{(\arcsec)}
& \colhead{Exposures}
& \colhead{(sec)}
& \colhead{Sky\tablenotemark{c}}
}
\startdata
\cutinhead{Pre-Perihelion}
2004 Jun 21&UH2.2 m& 3.80 & 3.21 & 13.61 &0.80 & $R\times2$ & 1200 & C\\

2004 Jun 22&UH2.2 m& 3.79 & 3.21 & 13.76 &0.80 & $R\times3$  & 300 & C\\

\cutinhead{Post-Perihelion}
2005 Oct 20&UH2.2 m& 1.76 & 1.74 & 33.02 &0.60 & $R\times11$ &  30 & P\\

2005 Oct 22&UH2.2 m&1.78 & 1.74 & 32.80 &0.76 & $V\times2$, $R\times3$&90&P\\
           &       &     &      &       &     & $I\times2$ \\

2005 Dec 02&Spitzer &2.14 & 1.94 & 28.19 &$\ldots$ & MIPS24$\times14$ 
& 10 & $\ldots$\\

2005 Dec 21&UH2.2 m&2.32 & 1.72 & 22.56 &0.71&$R\times13$, $I\times2$&200&P\\

2005 Dec 22&UH2.2 m&2.32 & 1.72 & 22.32 &0.65&$V\times3$, $R\times18$&150&P\\
           &       &     &      &       &    & $I\times3$ \\

2005 Dec 31&Spitzer &2.40 & 1.89 & 23.61 &$\ldots$&IRAC4.5$\times60$ 
& 30 & $\ldots$\\

2005 Dec 31&Spitzer &2.40 & 1.89 & 23.61 &$\ldots$&IRAC8.0$\times60$
& 30 & $\ldots$\\

2006 Mar 05&UH2.2 m&2.91 & 2.24 & 16.37 &0.84&$R\times5$, $I\times1$&500&C\\

2006 Mar 06&UH2.2 m&2.92 & 2.25 & 16.45 &0.73&$V\times1$, $R\times4$&400&P\\
           &       &     &      &       &    & $I\times1$ \\

\enddata

\tablenotetext{a}{Phase angle (Sun-Comet-Earth) or (Sun-Comet-Spitzer).}
\tablenotetext{b}{Seeing measured from stellar point sources.}
\tablenotetext{c}{Sky shows either cirrus (C) or photometric (P) conditions.}
\end{deluxetable}
\clearpage

%
%
\begin{deluxetable}{lcccccc}
\tablewidth{0pt}
\tablecaption{DERIVED OPTICAL PHOTOMETRIC PARAMETERS
\label{table:tb_mags}}
\tablecolumns{7}
\tablehead{
\colhead{Observation}\\
\colhead{Date}
& \colhead{$r_{h}$}
& \colhead{$R$-band\tablenotemark{a}}
& \colhead{$V-R$\tablenotemark{a}}
& \colhead{$R-I$\tablenotemark{a}}
& \colhead{$m(1,1,0)$}
& \colhead{$Af\rho$}\\
\colhead{(UT)}
& \colhead{(AU)}
& \colhead{(mag)}
& \colhead{(mag)}
& \colhead{(mag)}
& \colhead{(mag)}
& \colhead{(cm)}
}
\startdata
\cutinhead{Pre-Perihelion}
2004 Jun 21&3.80&21.89$\pm$0.04 &$\ldots$ &$\ldots$  &17.19 &4.80\\

2004 Jun 22&3.79&21.95$\pm$0.06 &$\ldots$ &$\ldots$  &17.24 &4.55\\

\cutinhead{Post-Perihelion}
2005 Oct 20&1.76&15.91$\pm$0.01 &$\ldots$ &$\ldots$ &12.82 &130.66\\

2005 Oct 22&1.78&16.05$\pm$0.01&0.53$\pm$0.01&0.49$\pm$0.01&12.89&124.47\\

2005 Dec 21&2.32 &17.05$\pm$0.01 &$\ldots$ &0.49$\pm$0.01 &13.74 &83.00\\

2005 Dec 22&2.32&17.07$\pm$0.01&0.52$\pm$0.01&0.48$\pm$0.01&13.76 &82.07\\

2006 Mar 05&2.91 &16.37$\pm$0.01 &$\ldots$ &0.48$\pm$0.01  &14.80 &39.06\\

2006 Mar 06&2.92 &18.48$\pm$0.01 &$\ldots$ &$\ldots$ & 14.61 & 46.50\\
\enddata
\tablenotetext{a}{Average color measured in a 3\arcsec \ circular aperture.}
\end{deluxetable}
\clearpage

%
%
\begin{deluxetable}{lcccc}
\tablewidth{0pt}
\tablecaption{DERIVED MID-INFRARED PHOTOMETRIC PARAMETERS of 21P
\label{table:ir-photom}}
\tablecolumns{5}
\tablehead{
\colhead{Instrument /}
&
&
& S.B. \\
\colhead{Wavelength}
& \colhead{$r_{h}$}
& \colhead{$F_\lambda$\tablenotemark{a}}
& \colhead{Slope, k\tablenotemark{b}}
& \colhead{$F_\lambda$\tablenotemark{c}}
\\
\colhead{(\micron)}
& \colhead{(AU)}
& \colhead{($\times10^{-20}$~W~cm$^{-2}$~\micron$^{-1}$)}
& 
& \colhead{($\times10^{-20}$~W~cm$^{-2}$~\micron$^{-1}$)}
}
\startdata
IRAC 4.5  & 2.40 & $1.63\pm0.05$ & $-1.10\pm0.07$ & $0.81\pm0.04$ \\

IRAC 8.0  & 2.40 & $9.54\pm0.20$ & $-1.01\pm0.06$ & $4.08\pm0.10$ \\

MIPS 24.0 & 2.14 & $20.6\pm0.8$  & $-0.99\pm0.03$ & $6.47\pm0.26$ \\

\enddata

\tablenotetext{a}{$F_\lambda$ measured in a 10\arcsec{} radius 
($\rho$) aperture.}

\tablenotetext{b}{The logarithmic slope of the azimuthally averaged,
  surface brightness profile.}

\tablenotetext{c}{$F_\lambda$ equivalent to a 3741~km radius
  aperture with the same observing circumstances as the 2005~Dec~22 UT
  optical observations (i.e., corrected for \textit{Spitzer}-comet
distances, dust temperature, and activity rate ($Af\rho$).}

\end{deluxetable}
\clearpage

%

\end{document}